\documentstyle[aps,prb,graphicx]{revtex}
\begin{document}
\bibliographystyle{mskbst}
\title{Superconducting Density of States from the Magnetic Penetration Depth
       of Electron-Doped Cuprates La$_{2-x}$Ce$_x$CuO$_{4-y}$ and
       Pr$_{2-x}$Ce$_x$CuO$_{4-y}$}

\author{John A. Skint, Mun-Seog Kim, and Thomas R. Lemberger}
\address{Department of Physics, Ohio State University,
         Columbus, OH 43210-1106}
\author{T. Greibe and M. Naito}
\address{NTT Basic Research Laboratories, 3-1 Morinosato Wakamiya,
         Atsugi-shi, Kanagawa 243, Japan}

\maketitle
\begin{abstract}
From measurements of the magnetic penetration depth,
$\lambda(T)$, from 1.6 K to $T_c$ in films of
electron-doped cuprates La$_{2-x}$Ce$_x$CuO$_{4-y}$ and
Pr$_{2-x}$Ce$_x$CuO$_{4-y}$ we obtain the normalized density of states,
$N_s(E)$ at $T=0$ by using a simple model. In this framework, the flat behavior of
$\lambda^{-2}(T)$ at low $T$ implies $N_s(E)$ is small, possibly
gapped, at low energies. The upward curvature in $\lambda^{-2}(T)$
near $T_c$ seen in overdoped films implies that superfluid comes from an anomalously
small energy band within
about $3k_BT_c$ of the Fermi surface.
\end{abstract}

\section{INTRODUCTION}
Pairing symmetry in the electron-doped cuprates
Ln$_{2-x}$Ce$_x$CuO$_{4-y}$ (Ln = Nd, Pr, or La) is controversial.
Phase-sensitive,\cite{tsuei2} angle resolved photoemission
spectroscopy,\cite{armitage1} and some penetration depth,
$\lambda(T)$, measurements \cite{kokales1,prozorov1} on nominally
optimally doped Pr$_{2-x}$Ce$_{x}$CuO$_{4-y}$ (PCCO) and
Nd$_{2-x}$Ce$_{x}$CuO$_{4-y}$ (NCCO) samples suggest
\textit{d}-wave pairing. Other penetration depth measurements
\cite{alff2,skinta1,skinta2,mskim_d1} and the
absence of a zero-bias conductance peak in tunneling measurements
\cite{kashiwaya1,alff1} favor \textit{s}-wave
superconductivity. Recent penetration depth measurements on PCCO
and La$_{2-x}$Ce$_{x}$CuO$_{4-y}$ (LCCO) films
\cite{skinta2} and tunneling measurements on PCCO films
\cite{biswas1} find evidence for a \textit{d}- to \textit{s}-wave
pairing transition near optimal doping.

In this work we focus on understanding the unusual upward
curvature that appears in $\lambda^{-2} (T)$ near $T_c$ in
overdoped films.\cite{skinta1,skinta2} These are the
films that show gapped behavior at low $T$, so they are especially
important to understand. To that end, we develop a new analysis
method which enables us to invert $\lambda^{-2} (T)$ to obtain the
normalized superconducting density of states, $N_s (E)$, over a
wide energy range. In this model, upward curvature means that
superconducting effects are confined to energies less than about
3$k_B T_c$. A fuller description, including data on many more
films, will appear elsewhere.\cite{skinta3}

\section{EXPERIMENTS}
Films were prepared by molecular-beam epitaxy (MBE) on 10 mm
$\times$ 10 mm $\times$ 0.35 mm (film La4) or 12.7 mm $\times$ 12.7 mm $\times$ 0.35 mm (films P1, P3,
P7, La2, La5, and La7) SrTiO$_{3}$ substrates as detailed
elsewhere.\cite{naito01,naito02,naito03,naito04,naito05} The same growth
procedures and parameters were used for all films of a given
compound. Table~\ref{table01} summarizes film properties. Films
P1, P3, and P7 (La2, La5, and La7) are the underdoped, optimally
doped, and overdoped PCCO (LCCO) films of Ref.
\onlinecite{skinta2}. Ce concentrations, $x$, are measured
by inductively coupled plasma spectroscopy to better than $\pm
0.005$. The films are highly \textit{c}-axis oriented, and their
\textit{ab}-plane resistivities, $\rho_{ab}(T)$, are
low.\cite{skinta2}

\begin{table}[h]

\caption{Properties of
PCCO and LCCO films. Ce doping, $x$, is known to better than $\pm
0.005$. PCCO films are 1000\AA\space thick, and LCCO films are
1250\AA\space thick. $T_{c}$ and $\Delta T_{c}$ are location and
full-width of peak in $\sigma_{1}$. Absolute uncertainty in film
thickness and $\lambda^{-2}(0)$ is $\pm$10\%. $\rho_{ab}$($T_c^+$)
is the \textit{ab}-plane resistivity just above $T_c$.}
\label{table01}
\begin{tabular}{c c c c c c c}
\hline Film & Material & $x$ & $T_c$ [K] & $\Delta T_c$ [K] &
$\lambda (0)$ [\AA] & $\rho_{ab}$($T_c^+$) [$\mu \Omega$cm] \\
\hline \hline
P1  & PCCO & 0.128 & 22.5 & 1.8 & 3100 & 40  \\
P3  & PCCO & 0.145 & 24.2 & 1.0 & 1800 & 19  \\
P7  & PCCO & 0.156 & 21.5 & 2.4 & 2000 & 18  \\
\hline
La2 & LCCO & 0.087 & 28.7 & 0.8 & 3200 & 67  \\
La4 & LCCO & 0.107 & 28.9 & 0.4 & 2300 & 37  \\
La5 & LCCO & 0.112 & 29.3 & 0.9 & 2500 & 33  \\
La7 & LCCO & 0.135 & 21.7 & 1.0 & 2300 & 15  \\
\hline
\end{tabular}
\end{table}

We measure $\lambda^{-2}(T)$ with a low frequency two-coil mutual
inductance technique described in detail
elsewhere.\cite{turneaure_1,turneaure_2} A film is centered between two small
coils, and a current at 50 kHz in one coil induces eddy currents
in the film. The second coil measures the attenuated magnetic
field. We have checked that the $ac$ field is too small to create
vortices, except near $T_c$ where $\lambda^{-2}(T)$ is much less
than 1\% of its value at $T = 0$. All of our conclusions are drawn
from data taken in the linear response regime.

The film's sheet conductivity, $\sigma (T)d = \sigma _{1}(T)d -
i\sigma_{2}(T)d$, with $d =$ film thickness, is deduced from the
measured mutual inductance. We define $T_c$ and $\Delta T_c$ to be
the temperature and full-width of the peak in $\sigma_1$. $\lambda
^{-2}(T)$ is obtained from $\sigma_2$: $\lambda ^{-2}(T) \equiv
\mu _{0} \omega \sigma _{2}(T)$ (MKS units). Experimental noise is
typically 0.2\% of $\lambda^{-2}(0)$ at low $T$ and is at least
partly due to drift in amplifier gain. The $\pm 10$\% absolute
uncertainty in $d$ is the largest source of error in $\lambda
^{-2}(T)$. This uncertainty does not impact the temperature
dependence of $\lambda ^{-2}(T)/\lambda ^{-2}(0)$. We estimate
film-to-film uncertainty in $\lambda^{-2}(0)$ to be $\pm 5$\%.
$T_c$'s determined from resistivity and penetration depth
measurements are identical.\cite{skinta1,skinta2}

\section{RESULTS}

\begin{figure}[htb]
\centerline{\includegraphics[height=2.5in]{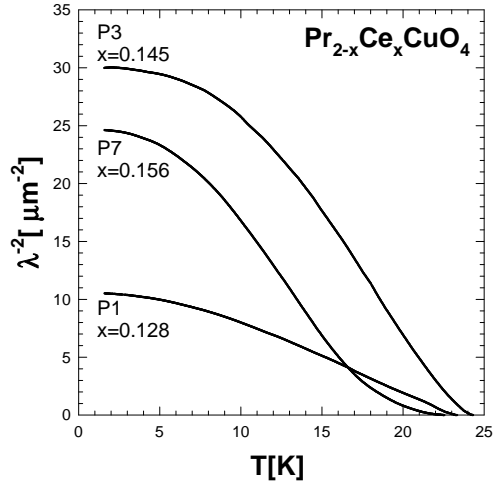}}
\caption{$\lambda^{-2}(T)$ in three Pr$_{2-x}$Ce$_x$CuO$_4$ films.
Film-to-film uncertainty in $\lambda^{-2}(0)$ is estimated to be
$\pm 5$\%.} \vspace{-0.1in} \label{ppd01-figure}
\end{figure}

Figure~\ref{ppd01-figure} displays $\lambda^{-2}(T)$ for PCCO
films.\cite{skinta2} Ref. \onlinecite{skinta2} also
shows data for LCCO. As detailed previously for films La2, La5,
La7, P1, P3, and P7,
\cite{skinta1,skinta2,skinta4} $\lambda^{-2}(T)$
at low $T$ is quadratic in $T$ at underdoping and exponentially
flat at overdoping.

Upward curvature in $\lambda^{-2}(T)$ appears near $T_c$ as seen
for films P3 and P7. Interestingly, the temperature range over
which upward curvature is observed does not increase monotonically
with doping. A study of the doping dependence of $\lambda^{-2}(T)$
in a series of nine Pr$_2$CuO$_4$-buffered PCCO films reveals a
similarly exaggerated foot at an intermediate overdoping
level.\cite{mskim_d1,skinta4} We emphasize that this upward
curvature \emph{is not} due to inhomogeneity.\cite{note01}
Instead, we believe that it proceeds from an anomalously small
order parameter, per the following.

\section{ANALYSIS}
Our model extracts the essential elements of superconductivity. It
starts with the clean-limit expression for the normalized superfluid
fraction, $\rho_s(T)$:\cite{tinkham1}
\begin{equation}
\rho_s(T) = 1 - 2\int_{0}^{\infty} dE \left(- \frac{\partial
f(E)}{\partial E} \right) N_s(E,T),
\label{rhodos-equation}
\end{equation}
where $N_s(E,T)$ is the normalized density of states, and $f(E)
\equiv 1/(1 + e^{E/k_B T})$ is the Fermi function. We justify the
clean-limit assumption by estimating the scattering rate from
measured resistivity just above $T_c$ and $\lambda^{-2} (0)$, as
discussed below. We invert $\lambda^{-2}(T)$ to get $N_s(E,0)$.
$N_s$ obtained this way is, in effect, the density of states for
only those states that contribute substantially to $\rho_s$, not
the full density of states. The situation is analogous to the
density of states determined by tunneling, which is weighted by
tunneling matrix elements. We assume $N_s(E,T)$ is a function of
only $x \equiv E/\Delta_0(T)$, so its $T$ dependence comes from
the ``order parameter'', $\Delta_0 (T)$. Theoretically, the
dependence of the normalized order parameter,
$\Delta_0(T/T_c)/\Delta_0(0)$, on $T/T_c$ is quantitatively is insensitive to details
of the superconducting state. It is about
the same for gapped \textit{s}-wave and gapless \textit{d}-wave
superconductors, as captured by the following respective approximations:
\begin{equation}
\Delta_0(T/T_c)/\Delta_0(0) \approx \sqrt{ \cos[\pi(0.04+
0.96T/T_c)^2/2]},
\label{sdelta-equation}
\end{equation}
\begin{equation}
\Delta_0(T/T_c)/\Delta_0(0) \approx \sqrt{1-(T/T_c)^3 },
\label{ddelta-equation}
\end{equation}
These two functions are very close, within 5\%. We assume the same behavior
for our films. Finally, we assume that
$N_s(x)$ has a generic form: small at low energies, peaked
at $E = \Delta_0(T)$, then unity. The abrupt drop from peak to unity accounts for
upward curvature in $\rho_s$. This very simple form produces
such good fits to our data that further refinement appears to be
of limited usefulness. Specifically, we assume:
\begin{equation}
N_s(x)= \left\{
\begin{array}{ll} A(x-\delta)^n & \textrm{for $\delta \leq x < 1-\epsilon$} \\
B & \textrm{for $1-\epsilon \leq x \leq 1+\epsilon$} \\
1 & \textrm{for $x > 1+\epsilon$} \\
\end{array} \right. \label{dos-equation}
\end{equation}
The width of the peak in $N_s$ is not a critical parameter. We
typically choose it to be 20\% of $\Delta_0$, ($i.e., \epsilon =
0.1$), in our fits. Our conclusions are insensitive to peak width
if it is less than about 40\%. $\delta$ is a number less
than $1-\epsilon$ that allows $N_s$ to have a minimum gap,
$\Delta_{min} = \delta \Delta_0$. $n$ and $A$ determine $N_s$ at
energies below its peak, so they are used in fitting the first 5\%
drop in $\rho_s$. The peak energy, $\Delta_0(0)$, is the only free
parameter for fitting the rest of $\rho_s (T)$. The height of the
peak, $B$, is determined by conservation of states.

\begin{figure}[htb]
\centerline{\includegraphics[height=2.8in]{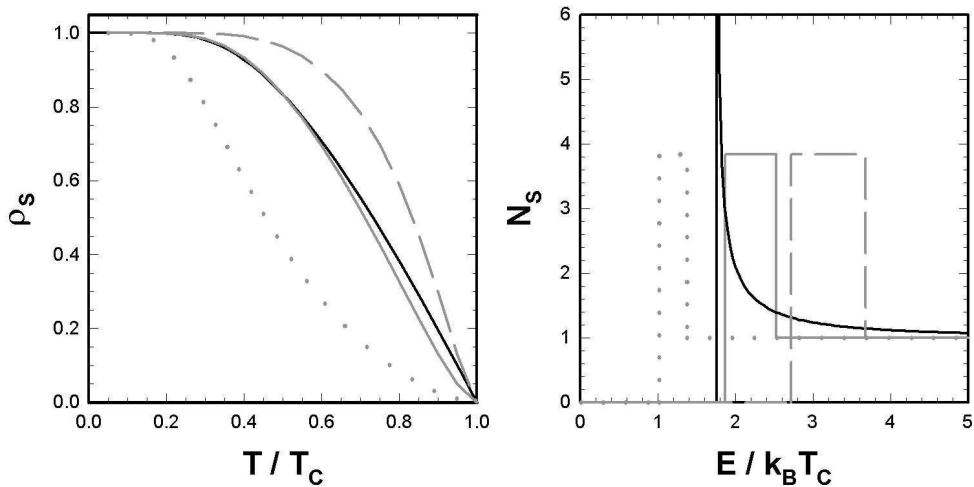}}
\caption{Theoretical superfluid fraction, $\rho_s(T/T_c)$, and
density of states, $N_s(E)$, (black lines, left and right panels)
for weak-coupling \textit{s}-wave superconductors. Gray lines are
approximate densities of states (right panel) and resulting fits
(left panel) calculated from the clean-limit model,
Eq.\ (\ref{rhodos-equation}). $\Delta_0(T/T_c)/\Delta_0(0)$ has the
\textit{s}-wave $T$-dependence of Eq.\ (\ref{sdelta-equation}).}
\vspace{-0.1in} \label{swaveb-figure}
\end{figure}

\begin{figure}[htb]
\centerline{\includegraphics[height=2.5in]{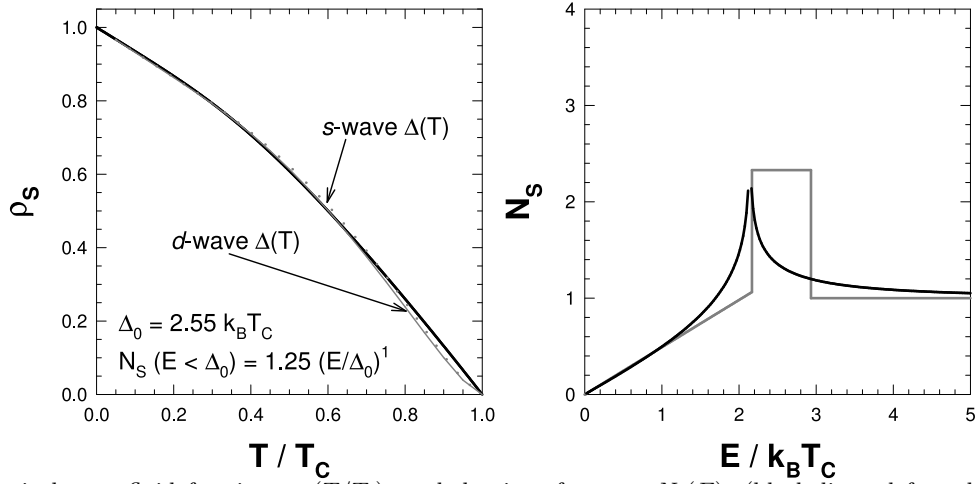}}
\caption{Theoretical superfluid fraction, $\rho_s(T/T_c)$, and
density of states, $N_s(E)$, (black lines, left and right panels)
for weak-coupling \textit{d}-wave superconductors. In the right
panel, the gray line is an approximate density of states with its
rectangular peak centered at $\Delta_0(0)=2.55$ $k_B T_c$. In the
left panel, the thin gray and dotted gray lines are the resulting
fits calculated using the approximate \textit{d}-wave
Eq.\ (\ref{ddelta-equation}) and \textit{s}-wave
Eq.\ (\ref{sdelta-equation}) gap $T$-dependences, respectively.}
\vspace{-0.1in} \label{dwaveb-figure}
\end{figure}

To get a feeling for the analysis, we apply it to obtain
approximate densities of states from calculated superfluid
fractions for weak-coupling \textit{s}-wave and \textit{d}-wave
superconductors, shown as black curves in the left-hand panels of
Figs.~\ref{swaveb-figure} and ~\ref{dwaveb-figure}. The ``fits'' (gray
curves) are calculated from the approximate densities of states
in the right-hand panels. In both cases, a good fit is
obtained when the low-energy edge of the peak in $N_s$ is close to
the peak in the exact density of states. To get a sense of how
accurately we can locate the peak in $N_s$, we calculated
$\rho_s(T)$ for three different peak positions, shown in
Fig.~\ref{swaveb-figure}. We emphasize that 
upward curvature in the $\rho_s$ fits near $T_c$ grows as the energy of
upper edge of the peak in $N_s$ decreases. Upward curvature is most obvious
when superconductivity does not extend to energies above about
$3k_B T_c$. In this case, states that contribute to superfluid density
are accessible at lower temperatures than in the true density of
states, which decreases slowly to unity at high energies, 
and the approximate $\rho_s(T)$ drops below the true
$\rho_s(T)$ at high temperatures.

For the \textit{s}-wave case, the \textit{s}-wave gap
approximation of Eq.\ (\ref{sdelta-equation}) was used in the model
calculations. $\Delta_{peak} \equiv (1-\epsilon)\Delta_0(0)$ is
defined as the low-energy edge of the peak in $N_s$. For this
\textit{s}-wave ``test case'' $\Delta_{peak} = 1.87$ $k_B T_c$,
only 6\% above the true weak-coupling gap value. For the
\textit{d}-wave case, both the approximate \textit{s}-wave
Eq.\ (\ref{sdelta-equation}) and \textit{d}-wave
Eq.\ (\ref{ddelta-equation}) gap $T$-dependences were used to
calculate $\rho_s(T)$ from the same approximate $N_s(E)$. Figure ~\ref{dwaveb-figure}
shows that there is little difference in quality between
the two $\rho_s(T)$ fits, demonstrating the insensitivity of this
analysis to the exact form of $\Delta_0(T/T_c)/\Delta_0(0)$.

We now turn to analysis of data. Results for one representative
film are presented here, with more results found elsewhere
\cite{skinta4,skinta3}. All $\rho_s$ curves in this section
are calculated using the approximate \textit{d}-wave gap
$T$-dependence of Eq.\ (\ref{ddelta-equation}), and with a peak width
of 20\% ($\epsilon=0.1$). ``Best fit'' in this section refers to the
approximate density of states that produces the most visually
pleasing match to the experimental superfluid fraction. Resulting
uncertainty in $\Delta_{peak} \equiv 0.9 \Delta_0(0)$ is estimated
to be $\pm 7.5$\%. Table~\ref{dos-table} provides a summary of
parameters from the analysis of several LCCO and PCCO films. For
all fits in Table~\ref{dos-table} $N_s$ is gapless ($\delta = 0$).

\begin{figure}[htb]
\centerline{\includegraphics[height=2.5in]{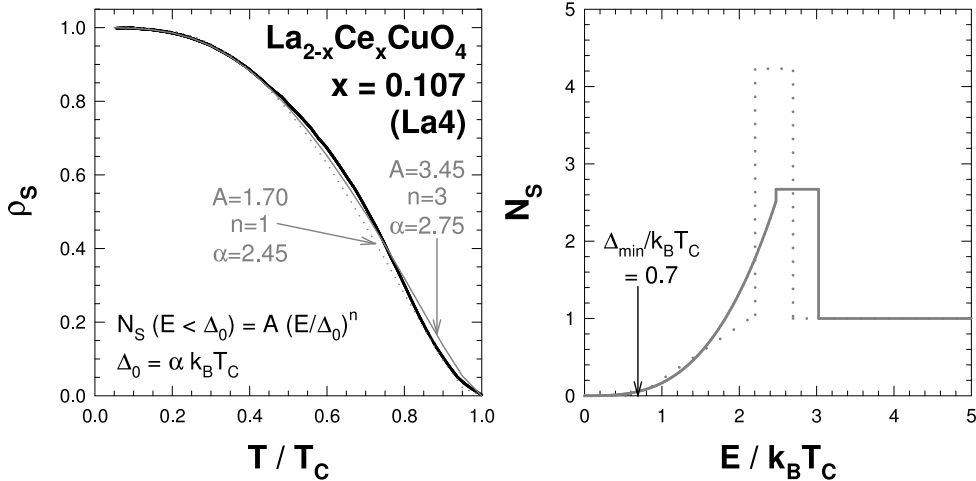}}
\caption{Experimental superfluid fraction in slightly overdoped
film La4 (thick black line, left panel). Solid gray lines are
$\rho_s(T)$ (left panel) and $N_s$ (right panel) for the case of
no gap. Dotted gray lines are $\rho_s(T)$ and $N_s$ for the case
of a minimum gap, $\Delta_{min}=0.7$ $k_BT_c$.} \vspace{-0.1in}
\label{la4dos2-figure}
\end{figure}

Figure~\ref{la4dos2-figure} shows data on film La4 (black curve)
and two acceptable fits, obtained from gapless (solid gray fit)
and gapped (dotted gray fit) densities of states. For the gapless
case, $N_s \propto (E/\Delta_0(T))^3$ at low energy, and for the
gapped case, $N_s$ vanishes for $E \leq 0.7k_BT_c$. Relative to
the gapless case, the peak in $N_s$ shifts down in energy to
compensate. At the level of accuracy of this analysis, we deem
that $\rho_s(T)$ fits produced by these two different densities of
states are both acceptable. $N_s$ may have a small gap, perhaps as big as $k_B T_c$. 
This is true for all films, even those that show quadratic behavior at low $T$.\cite{skinta3}

Fits of similar or better quality can be obtained for all of the
films that we have studied.\cite{mskim_d1,skinta3} This is
remarkable considering the simplicity of the model. We conclude
that upward curvature in $\lambda^{-2} (T)$ near $T_c$ indicates
that superfluid comes from states with energies less than about $3
k_B T_c$. Either superconductivity does not extend to higher
energies, or higher energy states do not contribute significantly
to superfluid density because their group velocity is small, or
there is a strong $k$-dependent scattering rate that diminishes
their current carrying ability.

\begin{table}[h]
\caption{Parameters from density-of-states analysis of
experimental superfluid density in Pr$_{2-x}$Ce$_x$CuO$_4$ and
La$_{2-x}$Ce$_x$CuO$_4$ films. $A$ and $n$ characterize $N_s$
below its peak: $N_s[E/\Delta_0(T)] = A[E/\Delta_0(T)]^n$ for $0 <
E/\Delta_0(T) < 0.9$. $\Delta_{peak} = 0.9 \Delta_0(0)$ is the
low-energy edge of the peak in $N_s$. Uncertainty in the best-fit
value of $\Delta_{peak}$ is $\approx \pm 7.5$\%.}
\label{dos-table}
\begin{center}
\begin{tabular}{c c c c c c}
\hline
Film & $x$ $(\pm 0.005)$  & $A$  & $n$ & $\Delta_0/k_BT_c$ & $\Delta_{peak}/k_BT_c$ \\
\hline \hline
P1   & 0.128 & 2.30 & 2   & 2.45              & 2.20     \\
P3   & 0.145 & 2.00 & 3   & 2.30              & 2.07     \\
P7   & 0.156 & 2.80 & 3   & 1.95              & 1.76     \\
\hline
La2  & 0.087 & 1.50 & 2   & 3.05              & 2.74     \\
La4  & 0.107 & 3.45 & 3   & 2.75              & 2.48     \\
La5  & 0.112 & 2.80 & 3   & 2.35              & 2.12     \\
La7  & 0.135 & 2.50 & 3   & 1.25              & 1.12     \\
\hline
\end{tabular}
\end{center}
\end{table}

We now return to our starting assumption that the films are clean,
in the sense that the electron mean free path is much larger than
the superconducting coherence length, or, in other words, that the
scattering rate, $\tau_{scatt}^{-1}$, is much smaller than the
order parameter, $\Delta_{peak}$. For present purposes, the usual
weak-coupling $s$-wave relationship among superfluid density,
$n_s(0)$, total carrier density, $n$, and $\tau_{scatt}^{-1}$
captures the decrease in $n_s (0)$ with scattering: $n_s (0)/n
\approx (1 + \hbar \tau_{scatt}^{-1}/ \pi \Delta_{peak})^{-1}$, so
samples are ``clean'' if $\gamma \equiv \hbar \tau_{scatt}^{-1} /
\pi \Delta_{peak}$ is small.\cite{tinkham1} $\tau_{scatt}^{-1}$ can
be estimated from the resistivity just above $T_c$ and
$\lambda(0)$:
\begin{equation}
\tau_{scatt}^{-1} \approx
\frac{\rho_{ab}(T_c^+)}{\mu_0 \lambda^{2}(0)}.
\label{scatt-equation}
\end{equation}
This approximation follows from the simple expressions for
penetration depth and resistivity: $\lambda^{-2}(0) =
n_{s}(0)e^{2}\mu_{0}/m \approx ne^{2}\mu_{0}/m$, and $\rho_{ab}
\approx m/n_{s}e^{2}\tau_{scatt}$. Using $\Delta_{peak}$ from Table~\ref{dos-table} and
$\tau_{scatt}^{-1}$ calculated from experimental resistivities and
penetration depths (Table~\ref{table01}), we find that $\gamma$ is
indeed small, varying between 0.17 and 0.23 for PCCO and 0.14 and
0.23 for LCCO.\cite{skinta4,skinta3}

It is interesting to explore the quantitative implications of a
small $\gamma$ in the contexts of \textit{s}-wave and
\textit{d}-wave superconductivity. In the context of
\textit{s}-wave theory, the deduced $\gamma$'s imply that
perfectly clean films would have superfluid densities
about 20\% larger than those reported here, a small correction. In
the context of dirty \textit{d}-wave theory,\cite{hirschfeld1} on
the other hand, the deduced $\gamma$'s imply that cleaner
films would have superfluid densities perhaps twice those reported
here, if scattering is in the unitary limit. That would mean that
$\lambda^{-2}(0)$ of optimally-doped PCCO and LCCO films would be about
double that of the closely-related hole-doped compound, LSCO.
\cite{panagopoulos1} This seems unreasonable to us, so we feel that if
superconductivity is \textit{d}-wave, scattering must be weak,
rather than unitary. But if so, then it is difficult to understand
why we do not observe the crossover from $T$ to $T^2$ at low $T$ predicted by
dirty \textit{d}-wave theory.

\section{CONCLUSION}

Our phenomenological analysis finds that upward curvature that
develops in $\lambda^{-2} (T)$ upon overdoping with Ce indicates
that the energy scale for superconductivity decreases anomalously.
For underdoped films, on the other hand, the peak in the
superconducting density of states appears at a reasonable energy,
about 2.5 $k_B T_c$. The reason for the decrease needs to be
found.

The effective superconducting density of states, $N_s(E)$, is low
at low energies for every doping level and could vanish for $E$
less than some minimum gap, $\Delta_{min}$. This behavior
conflicts with the behavior expected for clean $d$-wave
superconductors. Scattering rates deduced from $\rho_{ab}(T_c^+)
\lambda^{-2}(0)$ can account for the absence of a crossover from
quadratic to $T$-linear behavior, as predicted by dirty
\textit{d}-wave theory, but only if one accepts that the
superfluid density of very clean LCCO and PCCO are about twice
that of their hole-doped cousin, LSCO. We note that measurements
on PCCO films deposited onto a buffer layer instead of
directly onto SrTiO$_3$ display exponentially flat behavior at low
$T$ regardless of Ce concentration, consistent with a gapped density of
states.\cite{mskim_d1}

\section*{ACKNOWLEDGMENTS}
This work was supported in part by grant NSF-DMR 0203739.


\begin{thebibliography}{10}
\bibitem{note01}
We have observed upward curvature in $\lambda^{-2}$ near $T_c$ in many films,
  but never at temperatures below the fluctuation peak in $\sigma_1$. We
  ascribe upward curvature with non-zero $\sigma_1$ to film inhomogeneity.
\end{thebibliography}

\end{document}